\documentstyle[preprint,prb,aps,psfig]{revtex}

\begin{document}
\draft

\title{Laterally-doped heterostructures for III-N lasing devices}
\author{S. M. Komirenko and K. W. Kim}
\address{Department of Electrical and Computer Engineering, \\
North Carolina State University, Raleigh, North Carolina 27695-7911}
\author{V. A. Kochelap}
\address{
Institute of Semiconductor Physics, \\
National Academy of Sciences of Ukraine, Kiev, 252650, Ukraine}
\author{J. M. Zavada}
\address{U.S. Army Research Office, Research Triangle Park, NC 27709-2211}


\maketitle

\begin{abstract}
{To achieve a high-density electron-hole plasma in group-III nitrides for efficient 
light emission, we propose a planar two-dimensional (2D) $p-i-n$ structure 
that can be created in selectively-doped superlattices and quantum wells. 
The 2D $p-i-n$ structure is formed in the quantum well layers due to efficient 
activation of donors and acceptors in the laterally doped barriers. 
We show that strongly non-equilibrium 2D electron-hole plasma with density 
above $10^{12}\,cm^{-2}$ can be realized in the $i-$region of the laterally 
biased $p-i-n$ structure, enabling the formation of interband population inversion 
and stimulated emission from such a LAteral Current pumped Emitter (LACE). 
We suggest that implementation of the lateral $p-i-n$ structures provides an 
efficient way of utilizing potential-profile-enhanced doping of superlattices 
and quantum wells for electric pumping of nitride-based lasers. 
}
\end{abstract}

\vskip 2cm
Currently AlGaN and InAlGaN based heterostructures are of great interest
for short-wavelength optoelectronic devices, i.e., light-emitting diodes
(LEDs), laser diodes (LDs) and photodetectors, operating with radiation that 
covers the spectral range from green to deep-ultraviolet.~\cite {Nakamura1,Pearton,Foxon}
Development of light emitting devices in this spectral range is stimulated by a number 
of practically important tasks including highly-efficient lighting, 
high-density optical storage, stimulation of chemical processes, bio-medical applications, 
etc. Pumped by the electrical current, blue LDs and
green-, blue-LEDs have been realized in InGaN/InAlN structures and are
commercially available.~\cite{Nakamura} 
Efficient UV electroluminescence 
of AlGaN has also been observed recently.~\cite{Hirayama}

Further improvement in the performance of group-III nitride based
optoelectronics requires solution of an important problem - obtaining
high-density hole currents in bipolar structures. It is well established 
that the difficulties in achieving high hole concentrations mostly come from 
the deep energy levels of known acceptors.~\cite{Pearton,Foxon,Schubert,Oder}

It has been shown~\cite{Schubert,Kozodoy,Waldron} that in a p-doped 
superlattice (SL) 
the average hole concentration can be considerably enhanced due to 
high activation efficiency of in-barrier acceptors that supply the 
holes into quantum wells (QWs). The holes, however, are mostly confined 
inside the QWs where  
their concentrations can exceed $10^{13}\,cm^{-2}$.~\cite{Waldron} 
The potential barriers that separate the QWs can be as high as 100 meV to 400 meV.
These barriers hinder participation of the holes in {\em the vertical 
transport} typical for standard light-emitting devices. In this Letter, we
propose to utilize {\em the lateral transport} of holes and electrons 
confined in group-III nitride QWs and SLs for high-intensity light 
emission.~\cite{remark-1}

\begin{figure}[tbp]
\psfig{figure=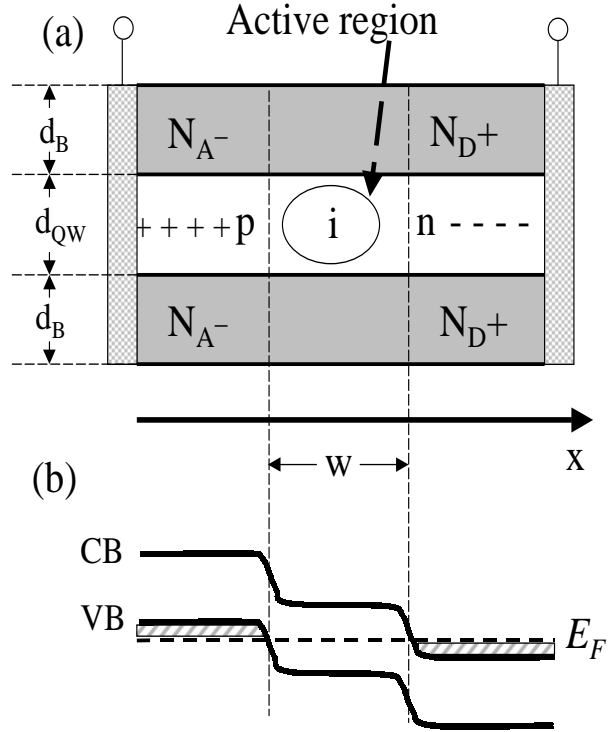,width=16.5cm,height=14.5cm,angle=0}
\vskip 5cm
\caption{ Schematic illustration of (a) the proposed multi-layered lateral 
$p-i-n$ structure and (b) energy band diagram in the lateral direction with 
no bias. CB and VB indicate the lowest populated energy levels in conduction and
valence bands, respectively.}
\label{f-1}
\end{figure}

The main element of the proposed structures is a single QW schematically 
illustrated in Fig.~\ref{f-1}. The QW layer is confined by 
selectively-doped barriers. Each barrier is doped laterally, so that 
an initial region doped with acceptors is followed by an undoped (intrinsic) region 
and, finally, by the region doped with donors. Thermal activation of 
the dopants in the barrier supplies carriers into the QW layer. Since the nitrides form type-I 
heterostructures, the QW layer accumulates both types of free carriers which 
lead to the formation of lateral $p-i-n$ structure.  Such $p-i-n$ structures can 
be fabricated by using re-growth techniques, 
position-dependent implantation methods,~\cite{Devis} etc. 
The contacts are to be made to $p$ and $n$ regions as shown in the
Fig.~1(a). 

The energy band diagram corresponding to an unbiased lateral $p-i-n$ structure is
shown in Fig.~\ref{f-1}(b). The energy barriers separating $p$, $i$, and $n$ 
regions arise due to the formation of charged regions.
Although the band bending is similar to that of a $p-i-n$ homostructure, 
the charge regions in a multilayered system are formed differently.
Indeed, for a $p-i-n$ homostructure the {\em local} charge neutrality
takes place in the $p$ and $n$ regions and the uniformly charged (depletion)
layers arise at the $p-i$ and $i-n$ junctions. In the planar multilayered structure,
the barrier and QW layers are electrically charged in the doped regions. The
quasi-neutrality occurs {\em on average} for any cross-section far from the
junctions. For the cross-sections near the junctions, the average charge is not
zero. This charge is responsible for the formation of potential barriers at
the junctions. In a forward  biased structure, the potential barriers decrease
providing for an injection of confined holes and electrons into the $i$-region.
This planar {\em double injection} gives rise to a non-equilibrium 
two-dimensional (2D) electron-hole plasma in the $i$ region. 
Radiative recombination of the plasma in the active region 
results in stimulated light emission. A SL arranged from QWs considered here 
will from an efficient souece of radiation - 
LAteral Current pumped Emitter (LACE). 

For the case of vertical transport, the theory of double injection in the 
$p-i-n$-homostructures and heterostructures is well developed.~\cite{Lampert}
It is based on a few well proven assumptions: (i) in electrically biased
structures, the $p,\,i$ and $n$ regions are mostly quasi-neutral, and (ii)
the charged (depleted) regions remain very narrow. The assumptions allow one
to avoid a detailed description of the processes in the depletion regions. To expand
this approach to lateral double injection, we need to analyze and compare the
length scales characterizing the structure under consideration. These
include different groups of the scales: the geometric scales - the QW layer
thickness $d_{QW}$, the barrier layer thickness $d_{B}$, and the size of the 
$i$ region $w$; the kinetic lengths - diffusion lengths of the electrons 
$L_{n}$, and holes $L_{p}$ (in general, different for the $p,\,n$ and $i$ 
regions); and the lengths of screening of an electric charge by carriers 
$l_{sc}^{n,p}$. In light emitting devices the $i$ region should be extended: 
$w\geq L_{n},L_{p}$, where the macroscopic diffusion lengths are of the order 
of ${\rm {\mu}m}$, 
while the screening lengths $l_{sc}$ are less than 10 nm.~\cite{Mitin} Thus,
we obtain the following inequalities: $w\geq
L_{n},L_{p}\gg d_{QW},d_{B}\geq l_{sc}\,.$

The lateral extensions of $p-i$ and $i-n$ junctions are estimated to be less than, 
or of the order of, $d_{QW},d_{B}$. Based on the above inequalities, we conclude that the
charged layers are very narrow and electron-hole recombination is negligible
there. These estimates allow one to study the lateral double injection by
the use of the Shockley approach.~\cite{Lampert} We can consider the extended 
$i$ region as quasi-neutral with the carrier
transport described by the bipolar drift-diffusion equations. Furthermore, 
the lateral electric fields inside the $p$ and $n$ regions are expected 
to be negligible (due to the uniformity along the x direction).  In the 
narrow charged regions, we subject the equations to the relevant boundary 
conditions.

In narrow QWs, the electrons and holes are
quantized and populate the lowest $n$ and $p$ subbands. Then, directing the 
$x$ axis along the QW, the equations for the bipolar plasma become:
\begin{eqnarray}  \label{eq-1}
&&\frac{d j_n}{d x} = \frac{d j_p}{d x} = - R\,, \\
&&j_n = - \mu_n n E - D_n \frac{d n}{d x}\,,\,\,\,\,\, j_p = \mu_p p E - D_p
\frac{d p}{d x}\,,  \label{eq-2}
\end{eqnarray}
where $E$ is the electric field; $n,\,p$ are the areal concentrations; $j_n,\,j_p$ 
are the flux densities; $\mu_n,\,\mu_p$ are the mobilities and $D_n,\,D_p$ are the 
diffusion coefficients for the electrons and the holes, respectively; and
$R$ is the rate of recombination. We will label these
parameters by the upper indices $p,\,i$ and $n$ for the respective 
regions. Equations~(\ref{eq-1}) and (\ref{eq-2}) 
have the same form 
for all regions. In the $i$ region, the quasi-neutrality condition yields:
\begin{equation}  \label{field}
E = \frac{J}{e (\mu_p^{(i)} + \mu_n^{(i)}) p^{{i}}} + \frac{D_p^{(i)}-
D_{n}^{(i)}}{(\mu_p^{(i)} + \mu_n^{(i)}) p^{(i)}} \frac{d p^{(i)}}{d x}\,,
\end{equation}
where $J$ is the electric current density calculated per QW. 
 The boundary conditions for Eqs.~(\ref{eq-1}) and (\ref{eq-2}) are: 
$p^{(p)}= p_0,\,n^{(p)}=0$ as $x \rightarrow - \infty$; 
$n^{(n)}=n_0,\,p^{(n)}=0$ as $x \rightarrow + \infty$. The fluxes of
minority carriers at the junctions are determined by the diffusion
processes: $j_n = - D_n^{(p)} \frac{d n^{(p)}}{d x}$ at $x=0$ and $j_p = -
D_p^{(n)} \frac{d p^{(n)}}{d x}$ at $x=w$. At the $p-i$ junction, the hole
concentration from the $i$ side [$p^{(i)}(0)$] and the electron
concentration from the $p$ side [$n^{(p)}(0)$] are determined by the energy
barrier $\Delta_p$ (of the $p-i$ junction) and the hole concentration $p_0$ 
in the $p$ side of the structure:
\begin{eqnarray}
&p^{(i)}(0)=&\frac{m_p k_B T}{\pi \hbar^2} ln \left[1+ e^{-\frac{\Delta_p}{%
k_B T}} \left(e^{\frac{\pi \hbar^2 p_0}{k_B T m_p}} - 1\right) \right]\,,
\label{bcs-1} \\
&n^{(p)}(0)=&\frac{m_n k_B T}{\pi \hbar^2} ln \left[1+ e^{-\frac{\Delta_p}{%
k_B T}} \left(e^{\frac{\pi \hbar^2 p^{(i)}(0)}{k_B T m_n}} - 1\right) \right]
\,.  \label{bcs-2}
\end{eqnarray}
Here, $\hbar$ is the Planck constant, $k_B$ is the Boltzmann constant, $T$ is the temperature, 
$m_n$ and $m_p$ are the effective masses of electrons and holes, respectively.  
Similarly, at the $i-n$ junction the electron concentration from the $i$ 
side [$n^{(i)} (w)$] and the hole concentration from the $n$ side 
[$ p^{(n)}(w)$] are determined by the energy barrier $\Delta_n$ (of the $i-n$ junction) 
and $n_0$. 
The voltage drop across the structure $V$ can be found as
\begin{equation}  \label{v-drop}
e V = E_G + E_{Fp}+ E_{Fn} - \Delta_p - \Delta_n - e \int_0^w E dx\,,
\end{equation}
where $E_G$ is the energy spacing between the electron and hole subbands in
the QWs, $E_{Fp}= k_B T \,ln \left[ exp\left(\frac{ \pi \hbar^2 p_0}{m_p k_B
T} \right) - 1 \right]$ is the Fermi level of the holes in the $p$ side of
the device, and $E_{Fn}$ has the same meaning for the electrons in the $n$ side
and can be calculated similarly. 

\begin{figure}[tbp]
\psfig{figure=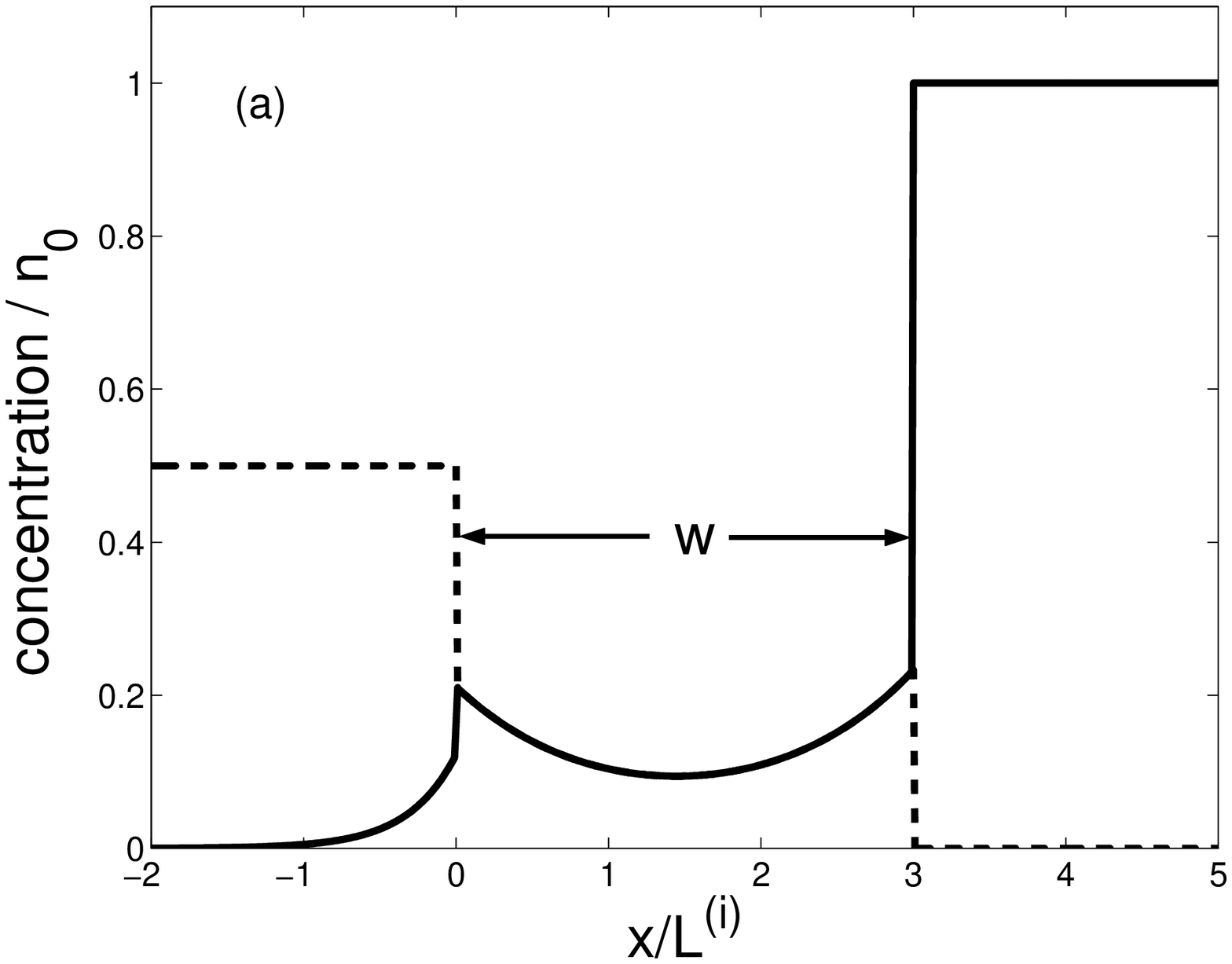,width=8.5cm,height=6.5cm,angle=0}
\psfig{figure=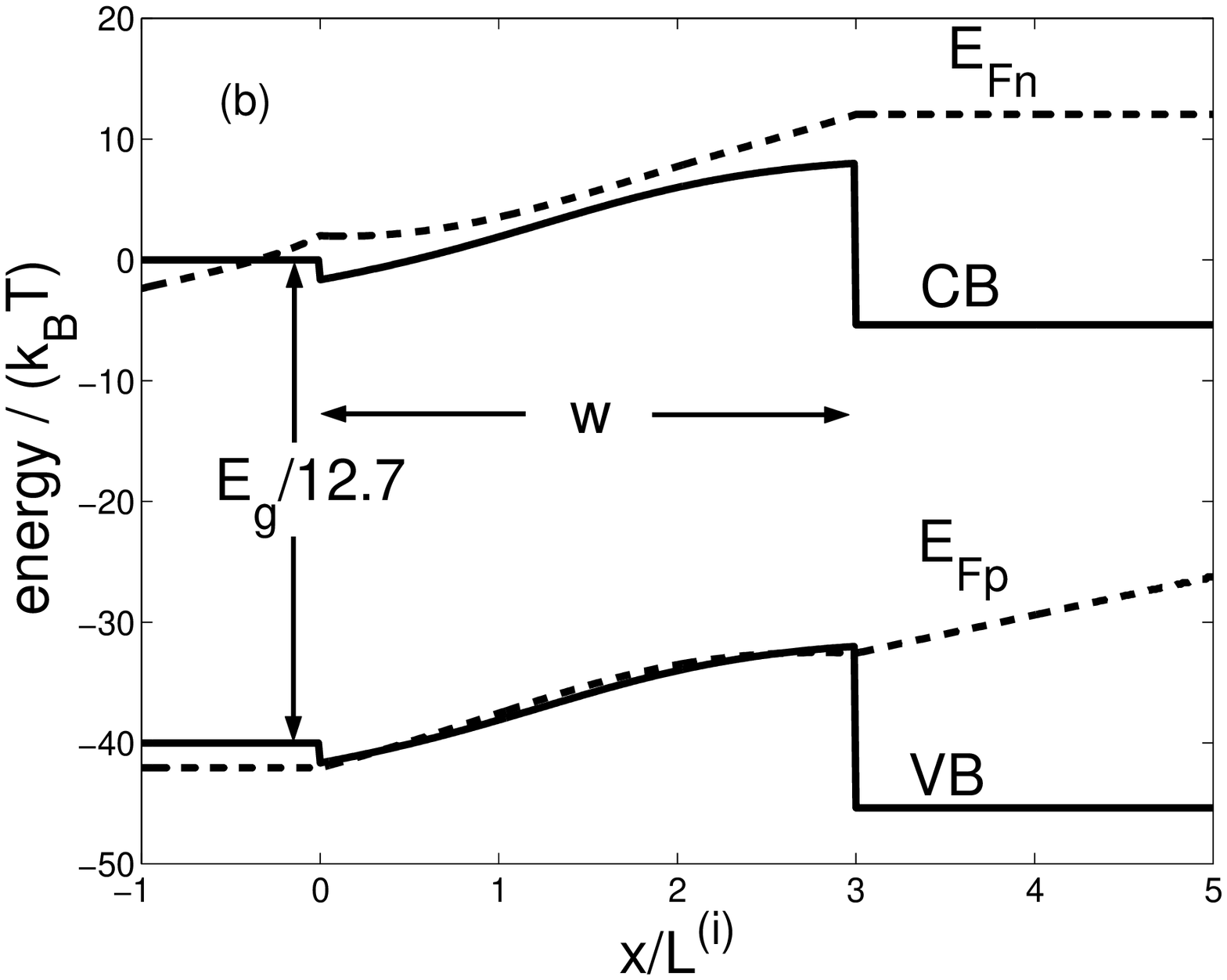,width=8.5cm,height=6.5cm,angle=0}
\vskip 1cm
\caption{Characteristics of the proposed lateral $p-i-n$ structure
with GaN QWs at $T=80 K$ in the double injection regime: (a) Concentration 
of the injected electrons and holes (in the unit of $n_0$) versus distance x
along the structure (in the unit of $L^{(i)}$ which is the diffusion length
in the $i$ region). The electron (hole) density is represented by the solid 
(dashed) line. (b) Energy band diagram of the structure. 
The energies are given in the unit of $k_B T = 6.9\,meV$. The depletion 
layer thickness is neglected as discussed in the text. The operating parameters are 
given in Table I [GaN (a)].
CB and VB indicate the lowest populated energy levels in conduction and
valence bands, respectively.}
\label{f-2} 
\end{figure}

For numerical estimates, we assume the linear recombination mechanism 
with the recombination time $\tau _{R}$ different in the different device 
regions. As  an example, let us consider a GaN/AlGaN LACE.
We set $\tau _{R}^{(p)}=\tau _{R}^{(n)}=0.1\,ns,\,
\tau _{R}^{(i)}=1\,ns$,~\cite{Foxon} $m_{n}=0.18\,m_{0},$ and $m_{p}=0.8\,m_{0}$, where 
$m_{0}$ is the free electron mass. Then we assume that
$\mu _{n}/\mu _{p}=20$ and $\mu _{n}=500\,cm^{2}/sV$,~\cite{remark-2} 
and that $D_{n}$ and $D_{p}$ can be estimated using the Einstein
relationship.~\cite{Mitin} For $T=80\,K$ this results in the ambipolar 
diffusion length $L^{(i)}=1.2\,\mu m$, and diffusion lengths 
$L_{n}^{(p)}=1.7\,\mu m$ and $L_{p}^{(n)}=0.4\,\mu m $. 
The carrier concentrations $p_{0},\,n_{0}$ are given in Table I.~\cite{} In 
Fig.~\ref{f-2}(a) we present the distribution of the electrons and 
holes injected into the $p-i-n$ structure. The data are
obtained for $w =3.6\,\mu m$ and $J=16\,mA/mm$. In the quasi-neutral region,  
the concentrations are nonmonotonic with a maximum $p_{M}$ at the $n-i$ 
junction and a minimum $p_{m}$ at the middle of this 
region. In the $p$ and $n$ regions, the minority carrier concentrations 
decay over the distances of about $L_{n}^{(p)}$ and $L_{p}^{(n)}$. 
The corresponding energy diagram is shown in Fig.~\ref{f-2}(b). 
The potential barriers at the junctions 
$\Delta _{p}$ and $\Delta _{n}$ are finite and decrease with increasing in 
current. For example, the $p-i$ junction barrier vanishes first at 
$J=188\,mA/mm$. The built-in potential related to the
diffusion field of Eq.~(\ref{field}) facilitates spreading of 'slow' holes 
through the extended $i$ region.  The total voltage drop across the $p-i-n$ 
structure with 5-$nm$ QWs is $4.22\,V$.

Data collected in Table I for GaN and InN QWs allow one to conclude 
that under the planar double injection, high densities of 
electron-hole plasma - above $10^{12}\,cm^{-2}$ - can be achieved. For 
5-$nm$ QWs, the radiative recombination of the plasma will produce 
light emission centered at $344\,nm$ and $587\,nm$ for GaN and InN QWs, 
respectively; the wavelength can be scaled readily to the deep-ultraviolet
range by using AlGaN or AlGaInN LACEs.  Our calculations show that under lateral 
injection, the interband {\em population inversion} 
occurs across the entire $i$ region for all cases presented in Table I. 
The population inversion can be reached at quite modest currents and biases. 
For instance, in a GaN-based LACE with ten QWs, a strip of area 
of $100 \times 3.6\,\mu m^2$ can be inverted in currents less 
than $16\,mA$ at $T=80\,K$ [Fig.~\ref{f-2}(b)].

In conclusion, contemporary group-III nitride technology allows 
fabrication  of novel structures for light emitters - laterally, 
selectively doped QWs and SLs. In such sructures, planar $p-i-n$ regions  
with high concentrations of 2D electrons and holes are formed and 
highly-efficient double injection occurs when a bias is applied along 
the QWs. This results in 
high densities of 2D electron-hole plasma in an extended $i$ region 
and population inversion of the conduction and valence bands. We 
suggest that planar double injection that occurs in LACEs 
is an efficient method for electrical pumping of short-wavelength 
nitride-based lasers.

This work was supported in part by  the US Army Research Office.

\newpage
Table~I. Parameters characterizing the lateral double injection in
the nitride-based $p-i-n$ structures with (a) $p_0 = n_0/2 =5 \times
10^{12}\,cm^{-2}$ and (b) $p_0 = n_0 = 10^{13}\,cm^{-2}$. U/e is the applied 
voltage. 
\vskip 0.2 cm
\begin{center}
\begin{tabular}{c|c|c|ccc|cc}
\hline\hline
QW & $T$ & J & $\Delta_p$ & $\Delta_n$ & U & $p_m $ & $p_M$ \\
Material & $K$ & mA/mm &  & $meV$ &  & $10^{12}$ & $cm^{-2}$ \\ \hline
\hline
GaN&&&&&&&\\
(a) & 80 & 16 & 11 & 92 & 66.6 & 0.9 & 2.3 \\
(b) & 250 & 80 & 20 & 42 & 284 & 2 & 6 \\ \hline
InN&&&&&&& \\
(a) & 80 & 37 & 12 & 153 & 74 & 1 & 2.9 \\
(b) & 300 & 110 & 27 & 73 & 285 & 2.4 & 6.6 
\end{tabular}
\end{center}

\newpage

\newpage

\end{document}